\documentstyle[psfig]{l-aa}
\def\deg{\ifmmode^{\circ}\;\else$^{\circ}\;$\fi} 

\begin{document}

   \thesaurus{06         
              (08.14.1;  
               08,16,6;  
               08,02,3;  
               08,05,3;  
               08,13.2)} 
 
   \title{PSR J2019+2425: A Unique Testing Ground for Binary Evolution} 

   \author{T. M. Tauris \inst{} 
          }

   \offprints{tauris@astro.uva.nl}

   \institute{Center for High-Energy Astrophysics, University of Amsterdam,
              Kruislaan 403, NL-1098 SJ Amsterdam, The Netherlands
             }

   \date{Received 19 December 1997 / Accepted 26 March 1998}

   \maketitle

   \begin{abstract}
If the theoretical relationship between white dwarf mass and orbital period
for wide-orbit binary radio pulsars is assumed to be correct, then the
neutron star mass of PSR J2019+2425 is shown to be $\sim 1.20 M_{\odot}$.
Hence the mass of the neutron star in this system prior to the mass
transfer phase is expected to have been $< 1.1 M_{\odot}$.
Alternatively this system descends from the accretion induced collapse (AIC)
of a massive white dwarf.\\ 
We estimate the magnetic inclination angles of all the observed wide-orbit
low-mass binary pulsars in the Galactic disk using the core-mass
period relation and assuming that the spin axis of an accreting neutron star
aligns with the orbital angular momentum vector in the recycling process
of the pulsar. The large estimated magnetic inclination angle of PSR J2019+2425,
in combination with its old age, gives for this system evidence against 
alignment of the magnetic field axis with the rotational spin axis.
However, in the majority of the similar systems the distribution of magnetic 
inclination angles is concentrated toward low values
(if the core-mass period relation is correct) and suggests that
alignment has taken place.
 
   \end{abstract}
 
A correlation between orbital period and companion white dwarf mass has
often been proposed to exist among low-mass binary pulsars (hereafter LMBPs).
However, it has been demonstrated (Tauris 1996) that observations of wide-orbit
LMBPs are difficult to fit onto the theoretical
relation proposed originally by Joss~et~al.~(1987).
In this letter we look at the consequences for the population of LMBPs under
the assumption that the theoretical relation is correct. 
For a review on the formation and evolution of binary millisecond pulsars,
see Bhattacharya \& van den Heuvel (1991).
 
\begin{table*}
\caption{Observed wide-orbit (class {A}) low-mass binary pulsars in the
         Galactic disk.}
\begin{flushleft}
\begin{tabular}{lccccccccccc}
\noalign{\smallskip}
\hline
\noalign{\smallskip}
      & & & & & \multicolumn{3}{c}{$M_{\rm NS}=1.4 M_{\odot}$} & 
      & \multicolumn{3}{c}{$M_{\rm NS}=1.8 M_{\odot}$}\\
\cline{6-8}\cline{10-12}
\noalign{\smallskip}
 PSR-name & $P_{\rm orb}$ & $P_{\rm spin}$ & $f$ & $M_{\rm WD}^{\rm PMc}$ &
$M_{\rm WD}^{i=60 \deg}$ & $M_{\rm WD}^{\rm min}$ & $i^{\rm PMc}$ & & 
$M_{\rm WD}^{i=60 \deg}$ & $M_{\rm WD}^{\rm min}$ & $i^{\rm PMc}$ \\
\noalign{\smallskip}
\hline
\noalign{\smallskip}
B0820+02 & 1232$^{\rm d}$ & 865 ms & 0.003 $M_{\odot}$ &
0.500 $M_{\odot}$ & 0.231 $M_{\odot}$ & 0.197 $M_{\odot}$ & 26.3$^{\deg}$ & &
0.271 $M_{\odot}$ & 0.232 $M_{\odot}$ & 30.2$^{\deg}$ \\
J1455-3330 & 76.2$^{\rm d}$ & 7.99 ms & 0.0063 $M_{\odot}$ &
0.305 $M_{\odot}$ & 0.304 $M_{\odot}$ & 0.259 $M_{\odot}$ & 59.9$^{\deg}$ & &
0.356 $M_{\odot}$ & 0.303 $M_{\odot}$ & 84.2$^{\deg}$ \\
J1640+2224 & 175$^{\rm d}$ & 3.16 ms & 0.0058 $M_{\odot}$ &
0.351 $M_{\odot}$ & 0.295 $M_{\odot}$ & 0.251 $M_{\odot}$ & 48.0$^{\deg}$ & &
0.345 $M_{\odot}$ & 0.294 $M_{\odot}$ & 58.5$^{\deg}$ \\
J1643-1224 & 147$^{\rm d}$ & 4.62 ms & 0.00078 $M_{\odot}$ &
0.341 $M_{\odot}$ & 0.142 $M_{\odot}$ & 0.122 $M_{\odot}$ & 23.0$^{\deg}$ & &
0.167 $M_{\odot}$ & 0.144 $M_{\odot}$ & 26.7$^{\deg}$ \\
J1713+0747 & 67.8$^{\rm d}$ & 4.57 ms & 0.0079 $M_{\odot}$ &
0.299 $M_{\odot}$ & 0.332 $M_{\odot}$ & 0.282 $M_{\odot}$ & 71.6$^{\deg}$ & &
0.388 $M_{\odot}$ & 0.330 $M_{\odot}$ & --- \\
J1803-2712 & 407$^{\rm d}$ & 334 ms & 0.0013 $M_{\odot}$ &
0.407 $M_{\odot}$ & 0.170 $M_{\odot}$ & 0.146 $M_{\odot}$ & 23.5$^{\deg}$ & &
0.200 $M_{\odot}$ & 0.172 $M_{\odot}$ & 27.1$^{\deg}$ \\
B1953+29 & 117$^{\rm d}$ & 6.13 ms & 0.0024 $M_{\odot}$ &
0.328 $M_{\odot}$ & 0.213 $M_{\odot}$ & 0.182 $M_{\odot}$ & 36.0$^{\deg}$ & &
0.250 $M_{\odot}$ & 0.214 $M_{\odot}$ & 42.5$^{\deg}$ \\
J2019+2425 & 76.5$^{\rm d}$ & 3.93 ms & 0.0107 $M_{\odot}$ &
0.305 $M_{\odot}$ & 0.373 $M_{\odot}$ & 0.316 $M_{\odot}$ & --- & &
0.435 $M_{\odot}$ & 0.369 $M_{\odot}$ & --- \\
J2033+1734 & 56.2$^{\rm d}$ & 5.94 ms & 0.0027 $M_{\odot}$ &
0.290 $M_{\odot}$ & 0.222 $M_{\odot}$ & 0.190 $M_{\odot}$ & 43.0$^{\deg}$ & &
0.261 $M_{\odot}$ & 0.223 $M_{\odot}$ & 51.8$^{\deg}$ \\
J2229+2643 & 93.0$^{\rm d}$ & 2.98 ms & 0.00084 $M_{\odot}$ &
0.315 $M_{\odot}$ & 0.146 $M_{\odot}$ & 0.125 $M_{\odot}$ & 25.4$^{\deg}$ & &
0.171 $M_{\odot}$ & 0.147 $M_{\odot}$ & 29.6$^{\deg}$ \\
\noalign{\smallskip}
\hline
\noalign{\smallskip}
\end{tabular}
\begin{list}{}{}
\item[] $M_{\rm WD}^{\rm PMc}\,$: mass of the white dwarf as expected from 
        the core-mass period relation -- cf. eq.(1). 
        We assumed $R_0=4950 R_{\odot}$.
\vspace{0.2cm}
\item[] $M_{\rm WD}^{i=60 \deg}$: mass of the white dwarf assuming an 
        inclination angle, $i=60 \deg\!$ of the binary system.\\
\vspace{0.2cm}
\item[] $M_{\rm WD}^{\rm min}$: mass of the white dwarf assuming an 
        inclination angle, $i=90 \deg\!$ of the binary system.\\
\vspace{0.2cm}
\item[] $i^{\rm PMc}\,$: orbital inclination angle of the system in order
        to obtain $M_{\rm WD}=M_{\rm WD}^{\rm PMc}$. 
\vspace{0.2cm}
\item[] A horizontal dash
        means that any inclination angle is inconsistent with such a high
        mass for the neutron star (i.e. $\sin i > 1$).
\end{list}
\end{flushleft}
\end{table*}

If the orbital period after the formation of a neutron star is relatively
large ($\ga$ a few days), then the subsequent mass transfer is driven by 
the interior nuclear evolution of the companion star after it evolved into a
(sub)giant and loss of orbital angular momentum by gravitational wave
radiation and/or magnetic braking can be neglected.
In this case we get a low-mass {X}-ray binary (LMXB) with a (sub)giant donor.
These systems have been studied by Webbink et al. (1983), Taam (1983) and
Joss et al. (1987). 
If mass is transfered from a less massive companion star 
to the more massive neutron star, the orbit expands 
and a stable mass transfer is achieved as the donor ascends the giant branch. 
Since the radius of such a donor star is a simple function of
the mass of the degenerate helium core, $M_{\rm core}$, and the Roche-lobe 
radius, $R_{\rm L}$, only depends on the masses and separation between
the two stars, it is clear that the final orbital period
($40^{\rm d} \la P_{\rm orb}^{\rm f} \la 1000^{\rm d}$) of the resulting 
binary will be a function of the final mass 
($0.20 \la M_{\rm WD}$/$M_{\odot} \la 0.45$)
of the helium white dwarf companion.

The relation between the orbital period of the recycled pulsar and
the mass of its white dwarf companion was recently re-derived by
Rappaport~et~al.~(1995) using refined stellar evolution calculations:
\begin{equation}
  P_{\rm orb} = 0.374 \left[ \frac{R_0 M_{\rm WD}^{4.5}}
                            {1+4M_{\rm WD}^{4}}+0.5 \right]^{3/2} 
                      M_{\rm WD}^{-1/2}
\end{equation}
where $P_{\rm orb}$ is given in units of days and $M_{\rm WD}$ is expressed 
in units of solar masses and $3300 < R_0/R_{\odot} < 5500$ is 
an adjustable constant which depends on the composition of the donor star
(the progenitor of the white dwarf).\\
In Table~1 we have compared observational data with the core-mass period 
relation given in eq.~(1) and derived the expected white dwarf mass and orbital
inclination angle in each of the 10 wide-orbit LMBPs in the Galactic disk,
assuming two different values for the neutron star mass, $M_{\rm NS}$,
and using the observed mass functions defined by:
\begin{equation}
  f\,(M_{\rm NS},M_{\rm WD}) = \frac {(M_{\rm WD} \sin i)^{3}}
                             {(M_{\rm NS} + M_{\rm WD})^{2}}
                        = \frac {4 \pi ^{2}}{G} \:
                          \frac {(a_{\rm p} \sin i)^{3}}{P_{\rm orb}^{2}}
\end{equation}
The recycling process is assumed to align the spin axis of the neutron star with
the orbital angular momentum vector as a result of $\sim 10^8$ yr of stable
disk accretion. Hence the orbital inclination angle, $i$, is equivalent
to (on average) the magnetic inclination angle,~$\alpha$, defined as the
angle between the spin axis and the center of the pulsar beam 
({\em viz.} line-of-sight).

Wide-orbit LMBPs form a distinct class of binary millisecond pulsars (class {A})
and are expected to have helium white dwarf companions -- cf. Tauris (1996).
For helium white dwarf companions in the interval 
$0.17 < M_{\rm WD}/M_{\odot} < 0.45$, we notice that the mass of the white dwarf
can be conveniently found from the following formula: 
\begin{equation}
   M_{\rm WD}=\left({\frac{550 P_{\rm orb}^4}{R_0^6}}\right)^{1/23}
\end{equation}
which is a simple fit to eq.~(1) with an error of less than 1\% in the 
entire mass interval, independent of $R_0$\footnote{For very small values 
of $R_0$ ($\sim 3300 R_{\odot}$) the above fit is only accurate to within 1\%
in the mass interval $0.20 < M_{\rm WD}/M_{\odot} < 0.45$.}.

By combining the above equations is is possible to calculate $M_{\rm NS}$
as a function of $i$ and $R_0$. For PSR J2019+2425
the mass of the neutron star is constrained to be remarkably low.
This is shown in Fig.~1.
  \begin{figure}
      \psfig{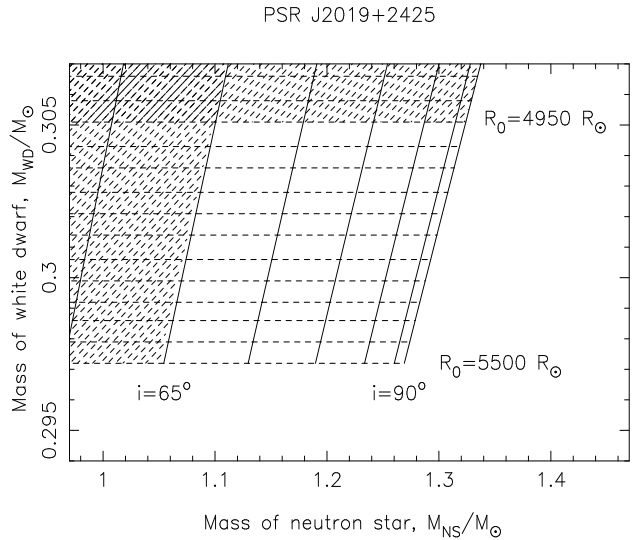}
    \caption{The expected mass of the pulsar PSR J2019+2425 
             as a function of $R_0$ (in units of $R_{\odot}$)
             and orbital inclination angle, $i$. 
             }
  \end{figure}
A weak interpulse is seen in the pulse profile of PSR~J2019+2425 
(Nice, Taylor \& Fruchter 1993) which indicates that $\alpha$, and hence $i$,
is large. Though such a pulsar profile could possibly be explained from a wide
one-pole emission beam (Manchester 1997), we shall assume 
$65\deg\!\!< i < 90\deg$.
We find that the value of $R_0$ is most likely to be in the interval:
$4950 < R_0/R_{\odot} < 5500$ -- i.e. the progenitor of the white dwarf
is either a pop.~I star or an ``intermediate'' pop.~I+II star,
cf. Rappaport~et~al.~(1995). Though the extremely large intrinsic
characteristic age, $\tau_{\rm i} = 27$~Gyr of PSR~J2019+2425 
(Camilo, Thorsett \& Kulkarni 1994) could suggest a progenitor star with
pop.~II abundances, we find it unlikely given the fact that the binary
is located in the Galactic disk ($|z|=100$ pc). 
If this is correct, it leaves us with a
neutron star mass of $M_{\rm NS} = 1.20\pm 0.10 M_{\odot}$ as our
best guess.

It has been suggested by van~den~Heuvel \& Bitzaraki (1995) that the
neutron star in PSR~J2029+2425 might have accreted as much as
0.65 $M_{\odot}$ in order to explain its present low magnetic field
strength, $B=1.8\times 10^8$ Gauss. However, in order to avoid a
pre-accretion neutron star mass of barely 
$M_{\rm NS}^{\rm pre-acc}=1.20 M_{\odot}-0.65 M_{\odot} \approx 0.6 M_{\odot}$ 
we suggest that the neutron star only has accreted $\sim 0.10 M_{\odot}$.
Another constraint on the maximum amount of matter accreted, and hence
on the minimum value of $M_{\rm NS}^{\rm pre-acc}$, is the fact that
this wide-orbit system is expected to have evolved though an {X}-ray phase
with stable Roche-lobe overflow and hence
$M_{\rm NS}^{\rm pre-acc} \ga M_2$ (where $M_2$ is the mass of the white
dwarf progenitor). However, we must require $M_2 > 1.1 M_{\odot}$, given
the large cooling age of 8--14 Gyr of this system (Hansen \& Phinney 1998),
in order for the companion to evolve in a time less than the age of our Galaxy
($\tau_{\rm MS} + \tau_{\rm cool} < \tau_{\rm gal}$). 
Therefore we also conclude
that the neutron star accreted less than 15\% of the transfered matter
($\Delta M = M_2-M_{\rm WD} > 0.80 M_{\odot}$) -- i.e. $\beta > 0.85$ 
(where $\beta$ is the fraction of the transfered matter lost from the system).
This is interesting since the mass loss rate of the donor star,
$\dot{M}_2$, in a system like PSR J2019+2425 is expected (Verbunt 1990) 
to have been less than the Eddington accretion limit,
$\dot{M}_{\rm Edd} \approx 1.5\times10^{-8} M_{\odot} {\rm yr}^{-1}$.

In Fig.~2 we have plotted the distribution of estimated magnetic inclination
angles, $\alpha$, from Table~1, assuming $M_{\rm NS}=1.4 M_{\odot}$
and $\alpha = i$ (see above).
There is seen to be a concentration toward low values of $\alpha$ in the
observed distribution. 
This is in agreement with the recent result obtained by Backer 1998 
(cf. his Fig.3) who analysed the distribution of observed minimum companion
masses\footnote{However, we disagree with his suggestion of a preference
for orthogonal magnetic and spin axes. Such configurations would also
exacerbate the already existing problem between the theoretical core-mass
period relation and observations of wide-orbit LMBPs (Tauris 1996).}.
Our result remains valid for other choices of $M_{\rm NS}$ and $R_0$
which only yield slight changes of the distribution. 
Since pulsars with small values of $\alpha$ generally shine on a smaller
fraction of the celestial sphere (simply due to geometry) it is clear
that the true underlying parent distribution is even further skewed
toward small values of $\alpha$. If the distribution of $\alpha$ 
(and thus $i$) was random, we would expect $\langle\alpha\rangle=60\deg$
for the parent population and $\langle\alpha\rangle > 60\deg$ for the
observed distribution.
Also keep in mind that systems where
$\alpha$ is smaller than the beam radius, $\rho$, can be very difficult to
detect due to lack of, or very little, modulation of the pulsed signal.
This could explain why no observed systems have $\alpha < 20\deg$.\\
  \begin{figure}
      \psfig{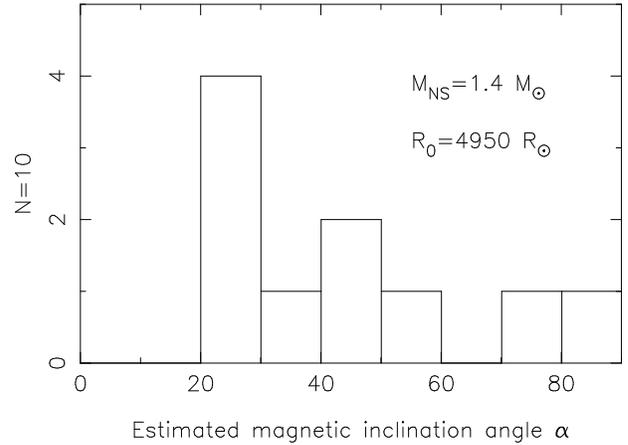}
    \caption{The distribution of estimated inclination angles in
             the wide-orbit LMBPs listed in Table~1, using the
             core-mass period relation. 
             }
  \end{figure}
For normal non-recycled pulsars there is no anti-correlation
between radio luminosity and $\alpha$ in a sample of 350 pulsars where
polarization studies have provided $\alpha$ (Tauris \& Manchester 1998).
Therefore there is no reason to believe that binary millisecond pulsars
are more easily detected when $\alpha$ is small.\\
Tauris \& Manchester (1998) have presented some evidence for alignment 
of the magnetic field axis with the spin axis of normal non-recycled
pulsars. Such a mechanism could also operate in recycled pulsars and
would be able to explain this non-random distribution of inclination
angles. 
This could also explain why $\tau_{\rm i}$ often exceeds the age of our Galaxy,
since alignment after the accretion process results in a braking index of
$n>3$ and therefore also in a deviation of $\tau_{\rm i}$ from the true age
(Manchester \& Taylor 1977).
However, in the case that alignment occurs in all recycled pulsars, 
PSR~J2019+2425 should be younger (due to its large value of $\alpha$) than the
bulk of the other wide-orbit LMBPs. This is in contradiction to the
very large cooling age of this system (see above) and the large value of 
$\tau_{\rm i}$ observed in this system compared
to that of the other systems -- although $\tau_{\rm i}$ is only a rough
age estimator individual to each system.
Alternatively, it is possible that there is an initial bifurcation angle above
which the (accretion) torque acting on the neutron star results in
a nearly perpendicular configuration after (or during) the mass transfer process
(van den Heuvel, private communication).\\
It should be noticed, that if alignment occurs in the majority of binary
millisecond pulsars this would enhance the birthrate problem between
LMXBs and LMBPs (Kulkarni \& Narayan 1988) since pulsars with smaller
magnetic inclination angles in average shine on a smaller part of the sky
and hence their Galactic population must be even larger.

An alternative model for the formation of PSR~J2029+2425 is that this
system descends from the accretion induced collapse
of a massive {O}-{Ne}-{Mg} white dwarf (e.g. Nomoto \& Kondo 1991).
In such a scenario the neutron star might have accreted only a very little
amount of matter after its formation and the orbital angular momemtum
axis need not be aligned with the spin axis of the neutron star.
Also eq.~(1) might not apply in this case
and thus we have no simple constraint on the lower limit to the
mass of the neutron star. 

Future observations of the shape and range of the general relativistic Shapiro 
delay in PSR~J2019+2425 would yield $i$ and $M_{\rm WD}$. The mass function 
would then give a value for the neutron star mass as well. These masses 
are highly desired in order to test theories for understanding the
formation and evolution of binary millisecond pulsars.

\begin{acknowledgements}
      The author thanks Ed van den Heuvel for discussions and many helpful 
      comments on the manuscript. 
      This research was supported in part by the NWO Spinoza-grant
      SPI 78-327.
\end{acknowledgements}

\end{document}